# A METRIC FOR THE ACTIVENESS OF A DISTRIBUTED OBJECT-ORIENTED COMPONENT LIBRARY

**Sachin Lakra[1], T.V. Prasad[2], Shree Harsh Atrey[3] and Deepak Kumar Sharma[1]**

[1]Department of Information Technology, Manav Rachna College of Engineering,
Faridabad, Haryana, India
[2]Department of Computer Science and Engineering, Lingaya's University,
Faridabad, Haryana, India
[3]Department of Computer Science and Engineering, Manav Rachna College of Engineering,
Faridabad, Haryana, India

*Abstract:* This paper makes an attempt to analyze the Activeness of a Distributed Object Oriented Component Library (DOOCL) and develops a software metric called Distributed Component Activeness Quotient(DCAQ) which is defined as the degree of readiness of a DOOCL. The advantages of the DCAQ include a possible comparison between various DOOCL's leading to selection of the best DOOCL for use during the development task, and providing a measure for gauging the usefulness of the DOOCL as indicated by the value of the DCAQ. The disadvantage of the DCAQ is that it may have some error because of its subjective and random nature.

The Stability of a DOOCL is another characteristic which is indicated by the DCAQ. The greater the value of the DCAQ, greater will be the stability of the corresponding DOOCL.

## 1. Introduction

Distributed Object Oriented Programming (DOOP) is based on the concept of components being assembled together to create a distributed software.

A programmer must necessarily meet project completion deadlines to avoid schedule and cost overruns, and achieve quality standards to avoid post-implementation difficulties. To support this, reusable software components have to be used and must be available to the programmer in the form of a component library, which, in turn, must be accessible to the programmer with from anywhere in the programming environment he is working on.

The paper examines the problem of the "*Activeness*" of a DOOCL and develops a software metric [4,6,7] called the *Distributed Component Activeness Quotient (DCAQ).* The DCAQ represents the readiness of a DOOCL to make components accessible to a programmer to help him achieve his two-fold goals of preventing schedule and cost overruns, and maintaining quality for stability [9] of the software.

The value of the DCAQ indicates the degree of organizedness, stability and responsiveness of a given DOOCL. These measurements indicate the improvements and enhancements required in the DOOCL. The values of the DCAQ for a number of DOOCL's allow an objective comparison between them, allowing a project manager to choose the best one for a given project.

## 2. Activeness

### 2.1 A Recent Concept.

Activeness is a recent new concept and is defined as the degree of readiness of a system to respond to the stimuli from the environment in which it exists [1]. Activeness may be related to any living entity or any system which exists in any environment in the universe. The Activeness of vacuum without any stimuli and no system existing in it is zero [1].

## 2.2 Why does Activeness exist in a system?

The question arises as to why a given system possesses Activeness. The answer is that every system has some "Organizedness" in it, that is, there is some degree of order in the system. The definition of a system itself says that "a System is a set of components working together to achieve a goal." Components cannot work together if they are not organized. This, in turn, implies that if a system is given an external stimulus there will be some change in the degree of order of the system, that is, there will be a response of the system to the stimulus. Every system responds to such stimuli. The "Organizedness" of the system makes it ready to respond to them to some degree. This degree of readiness of a system to respond to a stimulus is the concept called Activeness [3].

## 2.3 What is the need to study the Activeness of a system?

Another important question is that why Activeness should be studied at all. The answer lies in the fact that the observer wants to know how far she can depend on a system and its response. The system can respond well if it has the readiness to do so and if it is stable. It is stable if it is organized, i.e., if it has order. Thus if the Activeness of a system is known, its stability and how well it can respond to the stimulus it will be given, can be judged[3].

## 2.4 Origin of the Concept of Activeness

The author made the observation that if a human being tries to ward off a honeybee by waving a hand at it, the honeybee may get angered and may bite the human being. He wondered why the honeybee, being so small, is able to hurt a much bigger human being. The obvious answer was that the honeybee is *capable* of biting the human being and is therefore able to bite him. Further, the honeybee was capable because it was a system comprised of various biological subsystems, one of which was the sting, that is, the honeybee was very well organized and was ready to defend itself in case of an attack. This led to the idea of studying various systems as to how they can react to environmental stimuli.

## 3. Distributed Object Oriented Component Library (DOOCL) [2]

A Component Library is a repository of predefined software components which can be "picked off-the-shelf" from the Component Library and "placed" in a program. A Distributed Object Oriented Component Library (DOOCL) is an Object Oriented Component Library, which makes software components available to teams, working on modules of a distributed object oriented project, in a distributed manner [2]. The major advantage of such components is reusability. The reuse of components leads to the avoidance of reprogramming a particular task from scratch. Various sublibraries of a DOOCL must work in communion with each other, being accessible to all teams working on a Distributed Object Oriented Project (DOOP).

## 4. Working in a Distributed Object Oriented Environment (DOOE)

Distributed software runs on a permutation of a number of types of *machines* and a number of types of *operating systems.* The concept is borrowed from the idea of integrated circuits or chips, resistors, capacitors, etc., being brought together on a printed circuit board to form a piece of hardware. The components in hardware can be picked "off-the-shelf", placed on a printed circuit board and then soldered onto it. Similar is the case with DOOP where software components, i.e., predefined class definitions stored in a distributed object oriented component library (DOOCL) and its sub-libraries are "picked" and "placed" in a program. The "features" and "actions" of the software components, i.e., their properties and methods, respectively, are then used to implement the distributed software. The DOOCL's sublibraries which provide components to the programming environment may be geographically dispersed and accessible only through a private network or the Internet. This is the general approach with DOOP.

Programmers need to work very fast on a software project to meet project completion deadlines so as to prevent schedule overruns. Besides this, they have to maintain quality so that the software remains stable after completion and major changes are not required. For these two aspects, the software components must be readily available to the programmer in the form of a component library. The component library must be accessible to the programmer from anywhere in the DOOE. The smallest iota of a program is a single character typed by a programmer using the keyboard. Object Oriented Component Libraries in use today provide lists of classes, and the lists of their methods and their properties, at this level.

Work in a DOOE is done in an organized manner based upon the availability of components in the DOOCL being used.

## 4.1 The DolNet Distributed Object Oriented Software Process Model (DOOSPM)

The DolNet is a Division of Labour (Dol) based Distributed Object Oriented Software Process Model (DOOSPM). The generic approach taken in the DolNet involves the division of labour in a distributed manner among teams according to the types of *machines* and the types of *operating systems* that are going to run the DOO software. The work of a single team is related to a single combination of an operating system and a machine. The work of a single team is independent of the *number* of machines, if the machines are of the same *type*. Such a division of labour allows rapid application development to be applied. During coding, object oriented components are provided by a Distributed Object Oriented Component Library (DOOCL).

## 4.2 Steps involved in working in a DOOE using the DolNet DOOSPM

The steps involved in working in a DOOE using the Division of Labour based (DolNet) Distributed Object Oriented Software Process Model (DOOSPM) involves the following activities: [2]
- Integrated Study
- Integrated Analysis
- System Design
- Module Development Activity #n
- Integrated Testing
- Implementation & Maintenance

*Integrated Study:* The study of the DOO Project should be done in an integrated manner with the participation of at least one representative systems analyst from each team involved.

*Integrated Analysis:* The analysis should involve systems analysts and software engineers from each team working on the project with emphasis on understanding the system as an integrated whole.

*System Design:* The system design should take an overall view of the project and define the structure of the proposed system as a whole. This is the stage when work should be divided into modules based on the fact that the modules may be different either in their machine type and operating system combination or may be geographically dispersed forming a large project.

*Module Development Activity #n:* Each Module Development Activity for a given combination of a machine and an operating system, is further divided into the following sub-activities:

1. *Coding of programs*: This involves the use of one or more sublibraries of a DOOCL, and programs to be developed, by a separate team. During this development sub-activity, if a component is found to be available in a local or remote DOOCL sublibrary, the component is extracted from it. If a required component is unavailable in a local or remote sublibrary, it is developed as a new component and then added to the relevant sublibrary.
2. *Testing of Programs*: This sub-activity uses the programs produced in the previous sub-activity and a test suite of test cases.
3. *Module Testing Activity*: This is a constituent of a Module Development Activity but is conducted on the set of programs of the module as an integrated whole.

*Integrated Testing:* The testing activity consists of applying various integration testing techniques on the integrated software developed as a result of the various Module Development Activities.

*Implementation & Maintenance:* This activity involves installation and implementation of the various modules on their respective machine-OS combinations after completion of integrated testing and acceptance by the client(s)[2].

## 4.3. Problems in working in a Distributed Object Oriented Environment (DOOE)

The problems in working in a distributed object oriented environment with respect to accessing components from a DOOCL include:-
- If the DOOCL sublibrary does not contain the required component, the component may be available at another location, in which case the programmer has to search these other locations. The locations may be local, such as in another sublibrary, or remote, i.e., on a server in a network or at a website on the World Wide Web. Accessing such locations requires a large amount of time which leads to the deadline for the software project getting cumulatively delayed.
- The component may not be available at all in the DOOCL or at any other location, which leads to the need to develop this newly required component, again adding to the possibility of a schedule overrun.
- Synchronization and replication overheads to maintain upto date DOOCL sublibraries add to the access time delays.
- Poor network conditions can be damaging for a DOOP as both accessibility and access time degrade under such conditions.
- The DOOCL may not be organized well leading to the waste of time spent in searching for components in the Library.
- The non-availability of the component undermines the overall concept of reusability itself [1,2,3].

## 5. Factors involved in measuring the Activeness of a DOOCL.

There are five major factors upon which Activeness of a DOOCL is dependent. These include availability of components, access time of the component, the "Organizedness" of the DOOCL, the data rate of the network (in the case of a remote DOOE) and the number of sublibraries of the DOOCL.

### 5.1 Availability of Components.

The component is either available in the DOOCL or it is not available at all. If the component is not available then the Activeness of the DOOCL is simply 0 with respect to that component, otherwise it is 1.

### 5.2 Access Time of a Component.

Time in which the component can be accessed is dependent on the type of environment through which the component is made available to the programmer working at a workstation. The types of environments are:
- *Local environment* (on a desktop)- which may be:
  - Command Line Interface – e.g., JDK 1.3
  - Integrated Development Environment with a Graphical User Interface – e.g., VC++
- *Remote environments* which may be:-
  - A networked environment
  - The Internet

### 5.2.1 Access Time calculation in a Local Environment (on a desktop)

If the component is in a local environment, that is, on a desktop, then the access time may be calculated by taking the following components of the total time, starting from after the command has been issued by the programmer:

- Sending the command to the processor across the bus, given by

$$t_1 = \frac{b_c}{r_b} \quad (1)$$

where,
$b_c$ = no. of bits in the command,
$r_b$ = I/O bus transfer rate.

- Searching the memory by the processor for accessing the DOOCL sublibrary, given by

$$t_2 = hr \times t_c + (1-hr) \times (t_c + t_m) \quad (2)$$

where,
hr = hit ratio,
$t_c$ = time required to access cache,
$t_m$ = time required to access memory.

- Transferring the component to the video card for display, given by

$$t_3 = \frac{b_p}{r_b} \quad (3)$$

where,
$b_p$ = no. of bits in component,
$r_b$ = I/O bus transfer rate.

Thus, total time to access a component in a local environment is given by

$$T_l = t_1 + t_2 + t_3 \quad (4)$$

## 5.2.2 Access Time calculation in a Remote Environment (on a network)

If the component is in a remote environment, that is, on a network, then the access time may be calculated by taking the following components of the total time, starting from after the command has been issued by the programmer on his desktop:

- Sending the command to the processor across the bus, given by

$$t_1 = \frac{b_c}{r_b} \quad (5)$$

where,
$b_c$ = no. of bits in the command,
$r_b$ = I/O bus transfer rate.

- Sending the command across the network to the remote server which is storing the relevant DOOCL sublibrary, given by

$$t_2 = \frac{b_c}{r_n} \quad (6)$$

where,
$b_c$ = no. of bits in the command,
$r_n$ = data rate of the network.

- Searching the memory by the processor of the remote server for accessing the DOOCL sublibrary, given by

$$t_3 = hr \times t_c + (1 - hr) \times (t_c + t_m) \quad (7)$$

where,
hr = hit ratio,
$t_c$ = time required to access cache,
$t_m$ = time required to access main memory.

- Transferring the component to the network interface card within the remote server, for transfer across the network, given by

$$t_4 = \frac{b_p}{r_s} \quad (8)$$

where,
$b_p$ = no. of bits in component,
$r_s$ = I/O bus transfer rate of remote server.

- Transferring the component across the network to the programmer's computer, given by

$$t_5 = \frac{b_p}{r_n} \quad (9)$$

where,
$b_p$ = no. of bits in the component,
$r_n$ = data rate of the network.

- Transferring the component to the video card of the programmer's computer across the bus for display, given by

$$t_6 = \frac{b_p}{r_b} \quad (10)$$

where,
$b_p$ = no. of bits in component,
$r_c$ = I/O bus transfer rate at client.

Thus, total time to access a component in a remote environment is given by

$$T_r = t_1 + t_2 + t_3 + t_4 + t_5 + t_6 \quad (11)$$

The network conditions involved in the remotely accessible DOOCL sublibrary are a major factor in the success of the DOOCL. Network conditions are mainly dependent on the available data rate of the network.

### 5.3 "Organizedness" of a DOOCL.

The "organizedness" of a DOOCL is dependent on the type of organization of the DOOCL which may be:
- *Hierarchical Tree*
- *Sequential List*.

### 5.4 The number of sublibraries of the DOOCL.

The greater the number of sublibraries of the DOOCL, the greater will be the complexity of the DOOCL. Greater complexity will lead to slower responses from the DOOCL.

## 6. The Distributed Component Activeness Quotient (DCAQ).

### 6.1 Definition of the DCAQ.

**Definition 1.** *The DCAQ may be defined as the degree of readiness of a DOOCL to make a component available to a programmer and is given by*:

$$DCAQ = \frac{A_c \times T_s}{T \times n_s} \quad (12)$$

where  $A_c$ = Availability of the component in the DOOCL(can be 1 or 0),
$T_s$ = Organizedness of the DOOCL (Time taken to search the component according to the data structure used for storing the DOOCL and the time complexity of the search algorithm used to search the DOOCL for the required component),
T = Access time of the component, in seconds,
$n_s$ = Number of sublibraries of the DOOCL.

The DCAQ is dependent on the factors given in section 5. The value of $A_c$ may be 1 or 0 depending upon whether a component is available or it is not available. The value of $R_l$ depends upon the method followed by the DOOCL to make components accessible to the programmer. The faster the method, the greater is the weightage of the method and greater is the "organizedness" of the DOOCL. The data rate of the network determines the time of remote accessibility. The access time T may be either $T_l$ or $T_r$, as calculated in Equations 4 and 11, respectively. The number of sublibraries of a DOOCL is equal to the number of permutations of the number of types of machines and the number of types of operating systems.

### 6.2 Advantages

The advantages of the DCAQ include:
- Improvement required in the "Organizedness" of the DOOCL can be judged depending upon the value of the DCAQ.
- Comparison between DOOCL's can be made and the one with the maximum readiness may be chosen for a given project.
- Direct impact of a DOOCL on the schedule of the software project can be estimated numerically by using the DCAQ since the faster the programmer can access components, the faster the coding can be done and more the chances of meeting the project deadline.
- Stability of the software can be gauged as indicated by the value of the DCAQ since the software will be more stable if more components are available in the DOOCL and if more components can be reused, avoiding errors which can occur during rewriting of code. Higher the value of the DCAQ, more stable is the software [2,3].

### 6.3 Disadvantage

The DCAQ may have some error because it is subjective and stochastic in nature.

### 6.4 Applications of DCAQ

The DCAQ can be applied in the software industry wherever software development work is conducted in a DOOE to gauge the effectiveness of providing a DOOCL to a programmer.

## 7. Illustrations

**Illustration 1 :** Local Environment Scenario:

DOOCL resides in the Main Memory of a Desktop PC using a CLI and secluded from any network.
The DOOCL consists of 5 sublibraries.
The component required by programmer is C and it is known that C exists in the DOOCL.
The command for retrieving the component is retrcomp.
The component C consists of 16 lines of code, each line having 32 characters on an average, each character of 8 bits.

**Calculation of the DCAQ:**
**Availability** in DOOCL = 1
**Access Time** $T_l$:

Calculating $t_1$:
No. of bits in the command:
Suppose the command for retrieving the component is retrcomp, then
$b_c$ = 8 bits/character × 8 characters
= 64 bits
I/O bus transfer rate:
$r_b$ = 50 Mbps
= 0.05 bpns
Therefore,
$t_1$ = 64/0.05 = 1280 nsec

Calculating $t_2$:
Hit ratio, hr, of the cache = 0.9
$t_c$ = 20nsec
$t_m$ = 100nsec
Therefore,
$t_2$ = 0.9 × 20 + (1-0.9) × (20+100)
= 18 + 0.1 × 120
= 18 + 12
= 30 nsec

Calculating $t_3$:
Let C consist of 16 lines of code, each line having 32 characters on an average, each character of 8 bits. Then,
No. of bits in component C:
$b_p$ = 16 × 32 × 8
= 4096 bits
I/O bus transfer rate:
$r_b$ = 0.05 nbps
Therefore,
$t_3$ = 4096/0.05
= 81920 nsec

Calculating Access Time $T_l$ of C:
$T_l$ = 1280 + 30 + 81920 nsec
= 83230 nsec
= 83.23 microsec
= $8.323 \times 10^{-5}$ sec

**Organizedness** of the DOOCL:
Let the DOOCL be an alphabetically sorted Sequential List of Components
Let the Search algorithm used to search the DOOCL be binary search
Let the number of iterations, n, taken to successfully find C in the DOOCL = 16
Let the time taken for one iteration be 0.3 nsec
Then,
Time Complexity TC of successfully finding C using binary search will be
TC = $\log_2(n)$
= $\log_2(16)$ = 4
The time taken to successfully search C
= 4 × 0.3 nsec
= 1.2 nsec

Let the **number of sublibraries** $n_s$ in the DOOCL be 5.

Thus,
DCAQ = ( 1 × 1.2 ) / ( $8.323 \times 10^{-5}$ * 5 )
= 0.24 / $8.323 \times 10^{-5}$
≈ $0.028835 \times 10^5$
= $2.8835 \times 10^3$
= 2883.5

Indications based on the DCAQ:
1. Good organizedness.
2. High responsiveness.

**Illustration 2 :** Remote Environment

**Scenario:**
DOOCL resides in the Main Memory of a Server being accessed by a programmer from a client using a GUI over a Local Area Network secluded from the Internet. The DOOCL consists of 100 sublibraries.
The component required by programmer is C and it is known that C exists in the DOOCL.
The command for retrieving the component is retrievecompsrvr.
The component C consists of 32 lines of code, each line having 48 characters on an average, each character of 8 bits.

**Calculation of the DCAQ:**
**Availability** of C in DOOCL = 1
**Access Time** $T_r$ of C: Refer Table 1 for detailed calculations.

Calculating Access Time $T_r$ of C:
$T_r$ = 147217.2 nsec.
= 0.1472172 millisecond

= 1.472172 × 10⁻⁴ sec

Table 1. Total Access Time $T_r$ calculations for Illustration 2

| Time Component | No. of bits | | I/O Transfer Rate | | Data Transfer Rate of Network (bpns) | Hit Ratio | Access Time at server | | Time Component (ns) |
|---|---|---|---|---|---|---|---|---|---|
| | Command (bits) $b_c$ | DOOCL Component (bits) $b_p$ | Client (bpns) $r_c$ | Server (bpns) $r_s$ | $r_n$ | hr | Cache (ns) $t_c$ | Main Memory (ns) $t_m$ | |
| $t_1$ | 128 | | 0.05 | | | | | | 2560.0 |
| $t_2$ | 128 | | | | 0.1 | | | | 1280.0 |
| $t_3$ | | | | | | 0.92 | 10 | 90 | 17.2 |
| $t_4$ | | 4096 | | 0.2 | | | | | 20480.0 |
| $t_5$ | | 4096 | | | 0.1 | | | | 40960.0 |
| $t_6$ | | 4096 | 0.05 | | | | | | 81920.0 |
| **Total Access Time $T_r$** | | | | | | | | | **147217.2** |

**Organizedness**:
Data structure of the DOOCL :
  Alphabetically sorted binary tree
Search algorithm used :
  Binary search tree algorithm
Number of levels of the tree traversed to successfully find C in the DOOCL = 4
Let the time taken for traversing one level of the binary tree be 0.15 nsec
Then,
Time Complexity, TC, of successfully finding C using binary search tree algorithm will be
TC = $\log_2(n)$
   = $\log_2(4)$ = 2
The time taken to successfully search C
   = 2 × 0.15 nsec
   = 3 nsec

**Number of sublibraries** $n_s$ = 100

Thus,
DCAQ = ( 1 × 3 ) / (1.472172 × 10⁻⁴ × 100)
   = 3 / 0.01472172
   ≈ 203.7805

1. Average organizedness
2. Low responsiveness

## 8. Conclusion

Thus improvements in the DOOCL can be made based on the value of the DCAQ. The stability, organizedness, availability and responsiveness are features of the DOOCL which can be enhanced by taking numerical indications from the DCAQ.

## References

[1] Sachin Lakra, Bharti Jha, Nitin Bhardwaj, Ritu Saluja and Nand Kumar, "Metrics For The Pre-Development Phase Of Software Requirements Engineering"; Proceedings (Abstract) of National Conference on Emerging Trends in Software Engineering and Information Technology, Gwalior Engineering College, Gwalior, M.P., India; 29-30 March, 2007; pp. 21.

[2] Sachin Lakra and Deepak Kumar Sharma, "DolNet: A Division Of Labour Based Distributed Object Oriented Software Process Model"; Proceedings of International Conference on Data Management to be held at Institute of Management Technology, Ghaziabad, Uttar Pradesh, India; 25-26 February, 2008.

[3] Sachin Lakra, Nand Kumar, Sugandha Hooda, Nitin Bhardwaj; "A Metric For The Activeness Of An Object-Oriented Component Library", Proceedings of Software Engineering Research and Practice 2007 (SERP'07) of the WORLDCOMP'07 Conference, held at Las Vegas, Nevada, USA; 25-28 June, 2007; pp.704-709.

[4] Vaidya Dhananjay and Bhalerao Sidharth, "Overview of the existing Object Oriented Metrics and Frameworks"; Proceedings (Abstract) of National Conference on Emerging Trends in Software Engineering and Information

Technology, Gwalior Engineering College, Gwalior, M.P., India; 29-30 March, 2007, pp 15.

[5] Roger S. Pressman, "Software Engineering: A Practitioner's Approach", McGraw Hill, Sixth Edition, 2005.

[6] S. H. Kan, "Software Quality Engineering: Metrics and Models", Pearson Education, Asia, 2002.

[7] N. Fenton and Shari Pfleeger, "Software Metrics", Thomson Asia, Singapore, 2002.

[8] Grady Booch, *Object-Oriented Analysis and Design with Applications*, 2$^{nd}$ Edition, Addison-Wesley.

[9] http://www.pcmag.com/encyclopedia _term

[10] http://en.wikipedia.org/

[11] http://en.wikipedia.org/wiki/Distributed_ computing

[12] "IEEE Standard Glossary of Software Engineering Terminology," IEEE standard 610.12-1990, 1990.